\documentclass[fleqn, 12pt]{article}

\usepackage{authblk}
\title{Firearms Law and Fatal Police Shootings: \\ A Panel Data Analysis}
\author[1]{Marco Rogna\thanks{Correspondence: Marco Rogna, Department of Economics, Bochum University of Applied Sciences, Am Hochschulcampus 1, 44801 Bochum, Germany. Email: marco.rogna@hs-bochum.de}} 
\author[2]{Bich Diep Nguyen} 
\affil[1]{\footnotesize{Department of Economics, Bochum University of Applied Sciences}}
\affil[2]{\footnotesize{Faculty of Political Theory - Civic Education, Hanoi National University of Education}}
\date{}
\usepackage[a4paper, total={6in, 8in}]{geometry}
\usepackage{mathtools, bm, empheq}
\usepackage{natbib}
\usepackage{multirow}
\usepackage{url}
\usepackage{float, changepage}
\usepackage{subcaption}
\usepackage{chngcntr, changepage}
\usepackage[mathscr]{euscript}
\usepackage{filecontents}
\bibpunct{(}{)}{;}{a}{,}{,}
\usepackage[hidelinks]{hyperref}
\usepackage{booktabs, siunitx, pgfplotstable}
\usepackage{rotating}
\usepackage{graphicx}
\usepackage{endnotes}
\usepackage{enumitem}
\usepackage[mathscr]{euscript}
 \let\mathscr\relax
\usepackage{eurosym}
\usepackage[scr]{rsfso}
\usepackage{amsmath, amsthm, amssymb,tabu}
\usepackage{theoremref}
\usepackage[utf8]{inputenc}
\usepackage{setspace}
\usepackage{caption} 
\theoremstyle{definition}

\def\tb{\textbf}

\pgfplotsset{compat=1.14}
\DeclareSymbolFont{symbolsC}{U}{txsyc}{m}{n}
\DeclareMathSymbol{\notniFromTxfonts}{\mathrel}{symbolsC}{61}
\makeatletter
\newcommand{\customlabel}[2]{%
\protected@write \@auxout {}{\string \newlabel {#1}{{#2}{}}}}
\makeatother
\allowdisplaybreaks
\begin{document}
\let\texteuro\euro
\addtocontents{toc}{\protect\setstretch{1.5}}
\pagenumbering{gobble}
\maketitle
\newpage
\begin{abstract}
Among industrialized countries, U.S. holds two somehow inglorious records: the highest rate of fatal police shootings and the highest rate of deaths related to firearms. The latter has been associated with strong diffusion of firearms ownership largely due to loose legislation in several member states. The present paper investigates the relation between firearms legislation\textbackslash diffusion and the number of fatal police shooting episodes using a seven-year panel dataset. While our results confirm the negative impact of stricter firearms regulations found in previous cross-sectional studies, we find that the diffusion of guns ownership has no statistically significant effect. Furthermore, regulations pertaining to the sphere of gun owner accountability seem to be the most effective in reducing fatal police shootings.
\end{abstract}
\begin{adjustwidth}{1cm}{0cm}
\tb{Keywords:} Police shootings; Firearms law; Guns diffusion; Panel data.\\
\tb{J.E.L.} I12; I18.
\end{adjustwidth}   
\newpage
\pagenumbering{arabic}
\section*{Introduction}
The Unites States (U.S.) is one of the countries with the highest firearms--related death rate, exceptionally high compared to other industrialized countries \citep{pallin19}. Furthermore, it is the country with the highest firearms possession rate by citizens, thanks to loose legislation in several of its member states \citep{hemenway17}. The relation between firearms--related death rate and firearms possession rate has been largely investigated and a causal relation going from the latter to the former seems very plausible, e.g. \citet{krug98, bangalore13} and \citet{siegel13}. If this fact has prompted part of the public opinion to ask for stricter guns laws, the constitutional right to posses and bear firearms has been strenuously defended by the opposite faction, rendering this one of the most popular and controversial topic in the country.\par
Almost every aspect of the relation between firearms legislation and gun violence has been extensively researched. Besides the mentioned studies that have tried to establish a causal link between firearms--related death rate and firearms possession rate \citep{krug98, bangalore13, siegel13}, others have investigated the effect of firearms legislation\textbackslash possession rate on suicide rates \citep{kellermann92, anestis18}, on pediatric firearms--related mortality \cite{goyal19}, on homicides rate \citep{duggan01, kovandzic13, siegel13, siegel14}, and on the death rate of police officers on duty \citep{lester87, mustard01, swedler15}.\par
With approximately 1000 deaths per year, United States (U.S.) holds another inglorious record among industrialized countries: the highest rate of homicides committed by police forces \citep{hemenway19}. This raises the question of whether the high rate of police fatal shootings results from relaxed firearms legislation\textbackslash high possession rate. From a speculative point of view, one can argue that the diffusion of firearms increases the probability of police officers to face armed people while on duty, thus increasing the probability of being involved in potentially dangerous situations that require the use of guns from their part. Furthermore, the increased probability to face dangerous situations is a factor of stress that may lead law enforcement officers to overreact or, more generally, to commit mistakes. A positive relation between firearms diffusion and deadly assaults to police officers \citep{swedler15} corroborates the theory of an increase danger to officers for higher levels of firearms ownership.\par
Thanks to the availability of independent datasets that remedy the underreporting of police fatal use of force in official statistics \citep{conner19}, this topic has recently been investigated. \citet{hemenway19} find a positive association between firearms prevalence and fatal police shooting rates. \citet{kivisto17} report that U.S. states with stricter firearms legislation have lower incidence rates of fatal police shootings. Both studies are, however, cross-sectional and thus, despite the use of several controls common in the dedicated literature, may suffer from the omitted variable problem. \citet{kivisto17}, for example, mention, among the limitations of their paper, exactly this problem, stating that ``it is possible that states with stricter gun legislation also have better training for police officers and more stringent hiring practices, or that states that are already safe are more likely to implement stricter gun laws''. \par 
The novelty of the present study is to use a panel dataset to investigate the relation between firearms legislation\textbackslash possession rate and fatal police shootings. This allows to control for unobserved fixed characteristics at state level that may have biased previous analysis, providing, therefore, more robust results.
\section{Literature Review}
As discussed before, firearms legislation and ownership is a strongly investigated topic, particularly in the United States. A number of studies document negative impacts of firearms legislation and prevalence, such as increased suicide and homicide rate, although evidence is not unambiguous. From an extensive literature review, \citet{kleck15} concludes that guns diffusion is a positive determinant of crime rate, but this relation looses statistical significance in the most methodologically rigorous papers. \citet{branas09} find that possessing a gun increases the probability of being shot during an assault, thus dismantling the opinion of weapons having a protective role. On the other side, \citet{kleck93}, by comparing 170 U.S. cities, find scarce evidence that guns restrictions have some positive role in reducing the rate of violence. Similarly, \citet{altheimer08} evidences that gun availability has no effect in determining the number of total individual assaults and robbery, but only increases the number of the ones committed with a fire--weapon.\par 
Regarding one of the most serious violent crimes, homicide, the results are mixed. \citet{duggan01}, \citet{siegel13} and \citet{siegel14} find a significant and positive relation between guns diffusion and the number of homicides. However, this view is strongly opposed by \citet{kovandzic13}, which, on the contrary, document a negative relation. They cite a potential deterrent effect of guns as an explanation for their results. A similar debate has surrounded the permission of carrying concealed weapons, with \citet{lott97} showing a positive role of such law in reducing violent crimes whereas \citet{dezhbakhsh98} rejecting this finding and claiming the opposite. Findings regarding firearms diffusion and legislation are therefore very discordant even when considering aspects such as crime and homicide rate that are among the most studied.\par 
Shifting the attention on police forces, there is a paucity of research regarding the association between firearms diffusion and legislation and the number of killings of, and by, police officers. When considering the former, the killing of law enforcement officers, we have again contrasting findings. \citet{lester87} and \citet{swedler15} find that increasing levels of households gun ownership are a clear factor of risk for police officers, whereas \citet{mustard01}, limiting the attention to the possibility of carrying concealed weapons, puts in evidence a potential protective role of this law. \par
\citet{kivisto17} and \citet{hemenway19} investigate the relation between firearms and killings committed by police officers. The two studies both rely on independent and open source databases to retrieve the number of fatal police shootings: \citet{kivisto17} on \href{https://www.theguardian.com/us-news/series/counted-us-police-killings}{The Counted}, maintained by The Guardian, and \citet{hemenway19} on \href{https://www.washingtonpost.com/graphics/investigations/police-shootings-database/}{Fatal Force}, created by The Washington Post. Furthermore, they both rely on a cross sectional analysis using the fraction of suicides with a fire--weapon on the total number of suicides as a proxy for guns diffusion, as previously done in other papers, e.g. \citet{kleck04} and \citet{azrael04}. Compared to \citet{hemenway19}, whose focus is exclusively on guns diffusion as a cause of fatal police shootings, \citet{kivisto17} further consider firearms legislation, using the \href{https://www.bradyunited.org/}{Brady Campaign} scorecards as an indicator of law strength. In both papers, firearms ownership is found to positively affect the number of fatal police shootings. Even after controlling for firearms prevalence, firearms regulations on gun trafficking and on child and consumer safety significantly reduces fatal police shootings \citep{kivisto17}.\par 
Given the contrasting evidence emerged in other topics related to firearms diffusion and legislation and since both the last mentioned papers rely on a cross sectional analysis that may be plagued by the omitted variable bias, the extension to a panel data setting seems a necessary further step. This may help to strengthen the findings reached so far or to contest their validity as the result of a biased analysis.   
\section{Data and Methods}
Following \citet{kivisto17} and \citet{hemenway19}, our units of observation are the 50 U.S. states, with District of Columbia having being excluded for lack of data in several covariates. The covered time period is from January 1, 2012, to December 31, 2018, and all variables are expressed as yearly values, forming a dataset with seven time periods. Different databases have been consulted and merged in order to have all the variables of interest and the necessary controls: \hyperlink{https://fatalencounters.org/}{Fatal Encounters}, \href{https://giffords.org/lawcenter/resources/scorecard/}{Giffords scorecards}, the \href{https://www.census.gov/}{U.S. Census Bureau} data portal, the \href{https://crime-data-explorer.fr.cloud.gov/explorer/national/united-states/crime}{Federal Bureau of Investigation's (FBI's) Crime Data Explorer} and the Centers for Disease Control
and Prevention’s (CDC’s) \href{https://www.cdc.gov/injury/wisqars/index.html}{Web-based Injury Statistics Query and Reporting System} (WISQARS). Following is a description of all variables.
\subsection{Description of Variables}
Our dependent variable, the number of deaths caused by police shooting per million inhabitants (\textit{Pol\_Shoot}), is retrieved from the \hyperlink{https://fatalencounters.org/}{Fatal Encounters} database. We choose an independent database to alleviate the likely problem of underreporting of such episodes in the FBI's official statistics \citep{williams19}. Compared to other open source repositories, e.g. \href{https://www.theguardian.com/us-news/series/counted-us-police-killings}{The Counted}, \href{https://www.washingtonpost.com/graphics/investigations/police-shootings-database/}{Fatal Force}, \href{https://mappingpoliceviolence.org/}{Mapping Police Violence} and \href{https://www.gunviolencearchive.org/}{Gun Violence Archive}, the Fatal Encounters database covers the longest time span -- from 2000 to present. Specific cases of police shooting can be retrieved by selecting the category ``Deadly use of force'' and the subcategory ``Gunshot''.\par
Our independent variable of interest is the strength of firearms regulations at state level (\textit{Giff\_Score}). In order to obtain a synthetic measure of the strength of a state legislation, we rely on the \href{https://giffords.org/lawcenter/resources/scorecard/}{Giffords scorecards}, available for the period 2010-2018, with the exclusion of the year 2011, hence the need to drop the year 2010. The overall score is an aggregation of seven component scores, namely: background checks and access to firearms (\textit{BCAF}), other regulations of sales and transfers (\textit{ORST}), classes of weapons and magazines\textbackslash ammunitions (\textit{CWAM}), consumers and child safety (\textit{CCS}), gun owner accountability (\textit{GOA}), firearms in public places (\textit{FPP}) and a residual class (\textit{OTH}). Disaggregation of the overall score allows us to test the role of each component in explaining fatal police shootings, following \citet{kivisto17}. This is helpful in identifying the areas where intervention should be prioritized to reduce police shooting episodes. Since the scoring system has been slightly modified several times during the study period, we have implemented a harmonization procedure, retaining only the sub-indicators that remained unaltered over time. \citet{kivisto17}, using data from the same source, eliminate the weighting system in favor of a ``1 law = 1 point'' scale. They argue that a weighting system necessarily entails a degree of arbitrariness. However, we think that the equal weighting implied by the ``1 law = 1 point'' scale is analogously arbitrary. We, therefore, prefer to rely on the weights assigned by professional lawyers, thus leaving the Giffords scorecard scale unaltered.\par
Stricter legislation on firearms may reduce the quantity of fire--weapons owned by citizens, but may also promote safer use, e.g. by denying dangerous subjects access to guns or by increasing the safety of circulating weapons. Besides examining if laws to promote safe gun use are effective in reducing fatal police shootings, we can test if the effect of firearms legislations operate via the former channel by looking at the relation between fatal police shootings and gun diffusion. Lacking state--level data on gun ownership for our study period, we rely on a commonly adopted proxy (\textit{Suicide}) -- the percentage of suicides committed with a fire--weapon over the total of suicides \citep{kleck04, azrael04, kivisto17, hemenway19}. These data are retrieved from the Web-based Injury Statistics Query and Reporting System (WISQARS).\par 
\begin{table}[H]
	\centering
	\caption{Descriptive Statistics}\label{tab1}
	{\renewcommand{\arraystretch}{1.3}
		\begin{tabular}{lccccc}
			\hline\hline
			\textbf{Variable} & \textbf{N. Obs.} & \textbf{Mean} & \textbf{Std. Dev.} & \textbf{Min.} & \textbf{Max.} \\
			\hline\hline
			\textit{Pol\_Shoot} & 350 & 3.51 & 2.11 & 0 & 10.85 \\
			\textit{Giff\_Score} & 350 & 31.98 & 24.42 & 4 & 105.50 \\
			\textit{Crime} & 350 & 3627.32 & 1388.63 & 1026.24 & 8849.56 \\
			\textit{Suicide} & 350 & 51.54 & 12.35 & 13.20 & 74.30 \\
			\textit{PC\_Income} & 350 & 29712.58 & 4828.93 & 20119 & 52500.00 \\
			\textit{Urban} & 350 & 73.59 & 14.44 & 38.70 & 95 \\
			\textit{Poverty} & 350 & 9.99 & 2.80 & 4 & 19.20 \\
			\textit{White} & 350 & 76.95 & 12.67 & 24.30 & 95.10 \\
			\textit{Low\_Edu} & 350 & 11.19 & 2.97 & 6.10 & 18.60 \\
			\textit{Unemp.} & 350 & 3.96 & 1.18 & 1.80 & 7.90 \\
			\textit{Young} & 350 & 23.15 & 1.31 & 19.80 & 27.50 \\ \hline
			\multicolumn{6}{c}{\textit{\textbf{Giff\_Score disentangled}}} \\
			\textit{BCAF} & 350 & 7.19 & 5.51 & 0 & 22 \\
			\textit{ORST} & 350 & 4.12 & 5.89 & 0 & 24 \\
			\textit{CWAM} & 350 & 1.82 & 3.86 & 0 & 14 \\
			\textit{CCS} & 350 & 2.25 & 2.02 & 0 & 9 \\
			\textit{GOA} & 350 & 2.76 & 5.00 & 0 & 17.50 \\
			\textit{FPP} & 350 & 9.16 & 4.36 & 0 & 19\\ 
			\textit{OTH} & 350 & 4.67 & 2.18 & 0 & 10\\ \hline\hline
	\end{tabular}}
\end{table}
\indent 
Regarding control variables, we retrieved data on the number of violent crimes (per million inhabitants, \textit{Crime}) from the FBI's Crime Data Explorer, where a crime is defined as any of the four offenses -- murder and non-negligent manslaughter, rape, robbery, and aggravated assault. All the other controls are retrieved from the U.S. Census Bureau. Note that all values are projections on the 2010 census data. These controls include per--capita income in 2010 inflation--adjusted dollars (\textit{PC\_Income}) and the percentage of people living in urban areas (\textit{Urban}). It must be noted that, for this last variable, only figures for the year 2010 were available, thus it is treated as a time--invariant covariate. Other socio--economic characteristics, such as poverty rate (\textit{Poverty}),  unemployment rate (\textit{Unemp.}), and the percentage of adults with an education lower than high school diploma (\textit{Low\_Edu}), are also included. The percentage of young population, aged 18--34, (\textit{Young}) and the percentage of white Caucasians (\textit{White}) over the whole population are controlled for in the analysis. The last variable is added since several studies find a racial bias in police shootings \citep{ross15, nix17, mesic18}. Table \ref{tab1} reports some key statistics of all the mentioned variables.     
\subsection{Statistical Analysis}
The statistical analysis is divided into two main parts. In the first part, we focus on the role of the legislative strength as a whole, thus considering the overall score provided by Giffords for each U.S. state. In the second part, we analyze the component Giffords scores separately. This analysis should provide more specific policy indications with regard to the legislative field where intervention may be more productive in reducing fatal police shootings.\par 
The effects of any changer in legislation may take time to be observed. In all our analysis, therefore, both the overall and the component Giffords scores enter in their first lags. The inclusion of lags more distant in time (two or three years) is precluded by the limited number of time periods at our disposal. We have actually run regressions with the contemporaneous level of the Giffords scores, but none of them has resulted in significant coefficients (results are available from the authors upon request). The possibility to include both the lag and the current level is prevented by their high correlation ($\rho = 0.99$) that most likely causes a problem of collinearity. \par
\begin{figure}[htbp!]
	\centering
	\caption{Correlation Matrix}\label{fig1}
	\includegraphics[scale=0.75]{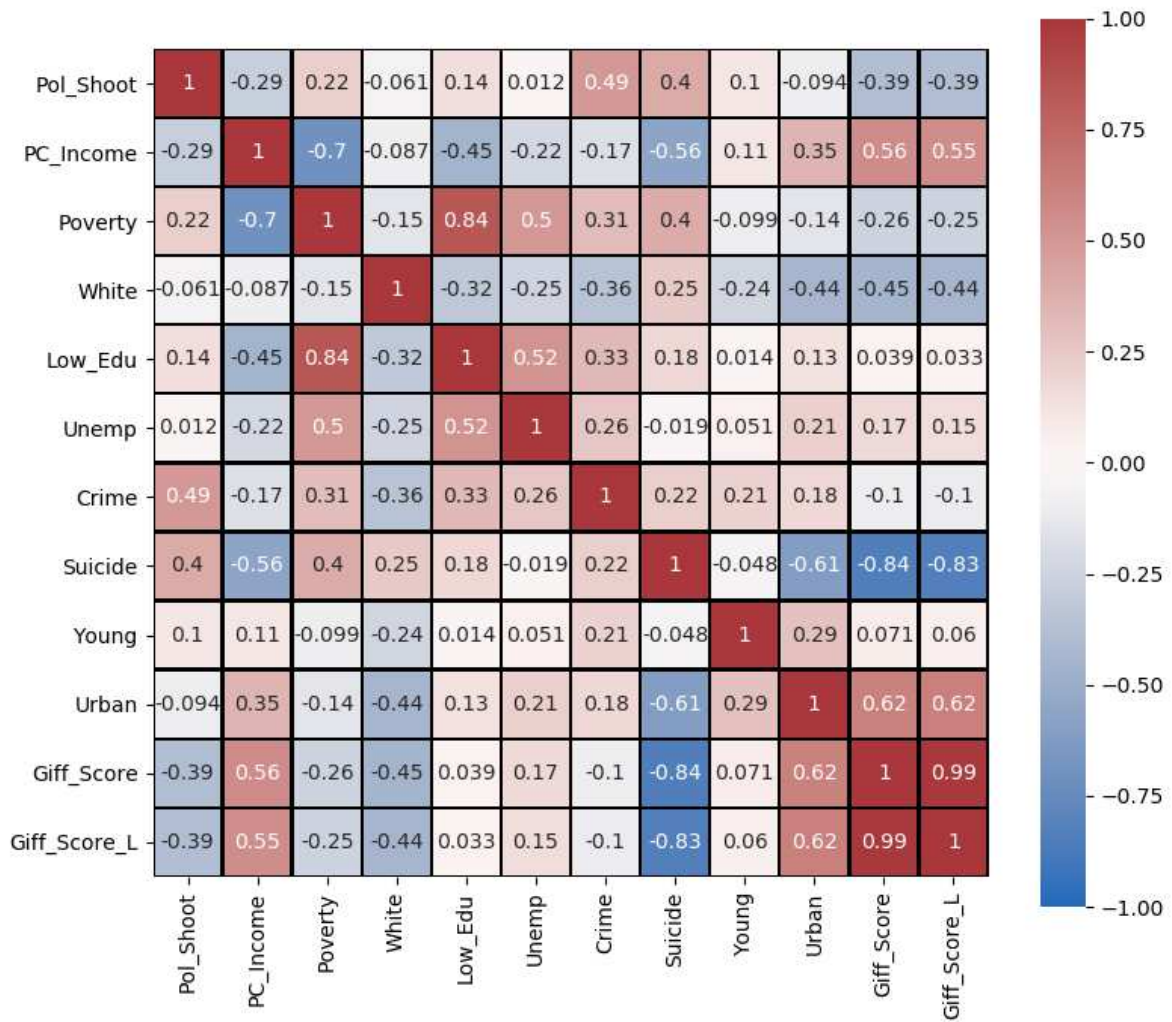}
\end{figure}

 
Figure \ref{fig1} shows the Spearman's rank correlation coefficients of all variables, except the component Giffords score. Note \textit{Giff\_Score\_L}  denotes the lagged Giffords score. From Figure \ref{fig1} it is possible to observe that several covariates have a relatively low correlation coefficient. The exceptions are the violent crime rate ($\rho = 0.49$) and the per--capita income ($\rho = -0.29$). \par 
\indent  
Despite the low correlation of these covariates with the dependent variable, our full specifications (specifications (1) and (2) in Table \ref{tab2}) include all of them. Their coefficients are not significant in these models. Failing to reject the null hypothesis that the joint significance of the socio--economic covariates (\textit{Poverty}, \textit{White}, \textit{Low\_Edu}, \textit{Unemp.} and \textit{Young}) is equal to zero (p--value = 0.6098 for model (1) and p--value = 0.8595 for model (2)), we omit these variables in any subsequent analysis. Despite its lack of significance, \textit{Urban} is kept in all models because of its high correlation with the lagged Giffords score ($\rho = 0.62$) and because it is a significant control in previous studies \citep{hemenway19}. Models (3) and (4) in Table  \ref{tab2} report the results with the parsimonious set of explanatory variables. From Table \ref{tabs1} in the Appendix, it is possible to observe non significant p--values for the RESET and for the Mundlak test. Compared to models (1) and (2), the coefficients of models (3) and (4) are negligibly different.\par
A conditional fixed effect (FE) and a random effect (RE) Poisson regressions (models (1) and (2) in Table \ref{tab2}) are our main models of interest. The functional form specification is checked through a RESET test, by adding the squared residuals of the Poisson regressions (FE and RE) and checking their significance \citep{ramsey74}. Results are reported in Table \ref{tabs1} in the  \hyperref[app]{Appendix} together with the coefficients of the year dummies (year 2013 as base) included in all models. The p--value of the squared residuals is far above the 10\% significance level, thus dismissing possible concerns about misspecification. The choice to report both the FE and RE estimates is due to the fact that the Mundlak test -- also reported in Table \ref{tabs1} -- has a significance level very close to the 5\% level \citep{mundlak78}. Although this test, chosen compared to the more common Hausman test given the presence of year dummies and a time invariant covariate (\textit{Urban}), suggests the use of the random effect model, the closeness of the p--value to the threshold level and the concerns about the distributional assumptions of the Poisson RE model lead us to report the fixed effect estimates as well. However, it can be observed that the estimated coefficients are very similar in both specifications.\par
\begin{table}[H]
	\centering
	\caption{Regressions Results: Total Giffords Score (Lagged)}\label{tab2}
	{\renewcommand{\arraystretch}{1.3}
		\begin{tabular}{lcccc}
			\hline\hline
			\multirow{3}{*}{} & \multirow{2}{*}{\textbf{\begin{tabular}[c]{@{}c@{}}(1) \\ Poisson FE\end{tabular}}} & \multirow{2}{*}{\textbf{\begin{tabular}[c]{@{}c@{}}(2) \\ Poisson RE\end{tabular}}} & \multirow{2}{*}{\textbf{\begin{tabular}[c]{@{}c@{}}(3) \\ Poisson FE\end{tabular}}} & \multirow{2}{*}{\textbf{\begin{tabular}[c]{@{}c@{}}(4) \\ Poisson RE\end{tabular}}} \\
			&  &  &  &  \\\hline\hline
			& \multicolumn{1}{l}{\textit{Pol\_Shoot}} & \multicolumn{1}{l}{\textit{Pol\_Shoot}} & \multicolumn{1}{l}{\textit{Pol\_Shoot}} & \multicolumn{1}{l}{\textit{Pol\_Shoot}} \\
			\multirow{2}{*}{\textit{Giff\_Score\_L}} & 0.990* & 0.992+ & 0.990* & 0.993+ \\
			& (-2.05) & (-1.65) & (-2.10) & (-1.82) \\
			\multirow{2}{*}{\textit{Crime}} & 1.000 & 1.000*** & 1.000 & 1.000*** \\
			& (1.61) & (4.04) & (1.46) & (4.69) \\
			\multirow{2}{*}{\textit{Suicide}} & 0.993 & 1.007 & 0.993 & 1.008 \\
			& (-0.77) & (1.03) & (-0.76) & (1.21) \\
			\multirow{2}{*}{\textit{PC\_Income}} & 1.000* & 1.000* & 1.000 & 1.000+ \\
			& (-2.15) & (-2.05) & (-1.49) & (-1.94) \\
			\multirow{2}{*}{\textit{Urban}} &  & 1.004 &  & 1.005 \\
			&  & (0.90) &  & (1.32)\\
			\multirow{2}{*}{\textit{Poverty}} & 0.958 & 0.965 &  &  \\
			& (-1.51) & (-1.04) &  &  \\
			\multirow{2}{*}{\textit{White}} & 0.958 & 0.996 &  &  \\
			& (-1.21) & (-0.66) &  &  \\
			\multirow{2}{*}{\textit{Low\_Edu}} & 1.025 & 1.012 &  &  \\
			& (0.90) & (0.38) &  &  \\
			\multirow{2}{*}{\textit{Unemp.}} & 1.064 & 1.032 &  &  \\
			& (1.20) & (0.80) &  &  \\
			\multirow{2}{*}{\textit{Young}} & 0.946 & 1.008 &  &  \\
			& (-1.01) & (0.23) &  &  \\ \hline\hline
			\multicolumn{5}{l}{\footnotesize{t-statistics in parentheses. P-value: + p \textless{}0.1, * p \textless{}0.05, ** p \textless{}0.01, *** p \textless{}0.001.}}
	\end{tabular}}
\end{table}
\indent
As a robustness check, we also report the results of linear models, both FE and RE, after that the dependent variable has been transformed with a Yeo--Johnson power transform in order to render its distribution more normal--like \citep{yeo00}. The results of the linear models are shown in Table \ref{tabs2}. Although the transformation of the dependent variable prevents the computation of meaningful marginal effects, the sign and significance of the coefficients serve to confirm the results of, or to signal a possible problem in, the Poisson regressions. The distribution of the dependent variable before and after the Yeo--Johnson transformation is shown in Figure \ref{figs1} in the \hyperref[app]{Appendix}.\par 
For our second purpose, we regress the rate of fatal police shootings on the component Giffords scores rather than the overall score. Here we present only the parsimonious models. Four models have been run, two Poisson -- FE and RE -- and two analogous linear regressions. Results are reported in Table \ref{tab3}, models (5), (6), (7) and (8), and auxiliary information can be found in Table \ref{tabs1} in the \hyperref[app]{Appendix}. From this last table, it is possible to see that the p--values of the RESET and of the Mundlak test are all above conventional significance levels. As for the previous models, the difference in the significance of the coefficients of the Poisson and of the linear regressions are very modest, so as the difference in the coefficients between the FE and RE models. \par 
\begin{table}[H]
	\centering
	\caption{Regressions Results: Linear Models}\label{tabs2}
	\resizebox{!}{9cm}{
		\begin{tabular}{lcccc}
			\hline\hline
			\multirow{2}{*}{} & \textbf{\begin{tabular}[c]{@{}c@{}}(1b) \\ Linear FE\end{tabular}} & \textbf{\begin{tabular}[c]{@{}c@{}}(2b) \\ Linear RE\end{tabular}} & \textbf{\begin{tabular}[c]{@{}c@{}}(3b) \\ Linear FE\end{tabular}} & \textbf{\begin{tabular}[c]{@{}c@{}}(4b) \\ Linear RE\end{tabular}} \\
			\hline\hline
			& \textit{Pol\_Shoot} & \textit{Pol\_Shoot} & \textit{Pol\_Shoot} & \textit{Pol\_Shoot} \\
			\multirow{2}{*}{\textit{Giff\_Score\_L}} & -0.00879* & -0.00924** & -0.00925* & -0.00803* \\
			& (-2.32) & (-2.86) & (-2.34) & (-2.52) \\
			\multirow{2}{*}{\textit{Crime}} & 1.47-E004+ & 1.36-E004*** & 1.60-E004+ & 149-E004*** \\
			& (1.76) & (3.75) & (1.95) & (4.43) \\
			\multirow{2}{*}{\textit{Suicide}} & -0.00515 & 0.00193 & -0.00562 & 0.00237 \\
			& (-0.55) & (0.36) & (-0.59) & (0.45) \\
			\multirow{2}{*}{\textit{PC\_Income}} & -1.83-E005* & -1.95-E005* & -1.44-E005+ & -155-E005* \\
			& (-2.07) & (-2.22) & (-1.75) & (-2.12) \\
			\multirow{2}{*}{\textit{Urban}} &  & 0.00255 &  & 0.00357 \\
			&  & (0.66) &  & (0.92) \\
			\multirow{2}{*}{\textit{Poverty}} & -0.0248 & -0.0259 &  &  \\
			& (-0.80) & (-0.87) &  &  \\
			\multirow{2}{*}{\textit{White}} & -0.0159 & -0.00411 &  &  \\
			& (-0.41) & (-1.03) &  &  \\
			\multirow{2}{*}{\textit{Low\_Edu}} & 0.0142 & 0.0124 &  &  \\
			& (0.51) & (0.52) &  &  \\
			\multirow{2}{*}{\textit{Unemp.}} & 0.0393 & 0.0327 &  &  \\
			& (0.64) & (0.69) &  &  \\
			\multirow{2}{*}{\textit{Young}} & -0.0476 & -0.0138 &  &  \\
			& (-0.92) & (-0.47) &  &  \\
			\multirow{2}{*}{\textit{Const.}} & 4.304 & 2.195* & 1.938** & 1.291* \\
			& (1.34) & (2.06) & (2.83) & (2.42) \\
			\multirow{2}{*}{\textit{2014}} & 0.0538 & 0.0639 & 0.0313 & 0.0418 \\
			& (0.76) & (0.92) & (0.46) & (0.60) \\
			\multirow{2}{*}{\textit{2015}} & 0.0661 & 0.0829 & 0.0359 & 0.0522 \\
			& (0.84) & (1.15) & (0.52) & (0.76) \\
			\multirow{2}{*}{\textit{2016}} & 0.137 & 0.151+ & 0.113* & 0.120* \\
			& (1.55) & (1.95) & (2.24) & (2.48) \\
			\multirow{2}{*}{\textit{2017}} & 0.163 & 0.189+ & 0.129+ & 0.146* \\
			& (1.42) & (1.75) & (1.99) & (2.31) \\
			\multirow{2}{*}{\textit{2018}} & 0.255* & 0.282* & 0.220** & 0.236*** \\
			& (2.11) & (2.40) & (3.10) & (3.36) \\ \hline\hline
			\multicolumn{5}{l}{\footnotesize{t-statistics in parentheses. P-value: + p \textless{}0.1, * p \textless{}0.05, ** p \textless{}0.01, *** p \textless{}0.001.}}
	\end{tabular}}
\end{table}
\indent

\begin{table}[H]
	\centering
	\caption{Regressions Results: Giffords Score Disentangled (Lagged)}\label{tab3}
	{\renewcommand{\arraystretch}{1.3}
		\begin{tabular}{lcccc}
			\hline\hline
			& \textbf{\begin{tabular}[c]{@{}c@{}}(5) \\ Poisson FE\end{tabular}} & \textbf{\begin{tabular}[c]{@{}c@{}}(6) \\ Poisson RE\end{tabular}} & \textbf{\begin{tabular}[c]{@{}c@{}}(7) \\ Linear FE\end{tabular}} & \textbf{\begin{tabular}[c]{@{}c@{}}(8) \\ Linear RE\end{tabular}} \\ \hline\hline
			& \textit{Pol\_Shoot} & \textit{Pol\_Shoot} & \textit{Pol\_Shoot} & \textit{Pol\_Shoot} \\
			\multirow{2}{*}{\textit{BCAF\_L}} & 1.001 & 1.000 & 0.00131 & -6.63-E004 \\
			& (0.15) & (0.03) & (0.22) & (-0.12) \\
			\multirow{2}{*}{\textit{ORST\_L}} & 1.032 & 1.008 & 0.0231 & 0.0110 \\
			& (1.44) & (0.47) & (1.17) & (0.93) \\
			\multirow{2}{*}{\textit{CWAM\_L}} & 0.981 & 1.014 & -0.0168 & 0.00372 \\
			& (-0.93) & (0.44) & (-1.01) & (0.22) \\
			\multirow{2}{*}{\textit{CCS\_L}} & 0.926 & 0.958 & -0.0732 & -0.0409 \\
			& (-1.09) & (-1.24) & (-1.19) & (-1.47) \\
			\multirow{2}{*}{\textit{GOA\_L}} & 0.963* & 0.962* & -0.0351* & -0.0352** \\
			& (-2.57) & (-2.32) & (-2.58) & (-2.93) \\
			\multirow{2}{*}{\textit{FPP\_L}} & 0.984 & 0.983 & -0.0191 & -0.0193 \\
			& (-1.25) & (-1.38) & (-1.54) & (-1.63) \\
			\multirow{2}{*}{\textit{Crime}} & 1.000 & 1.000*** & 1.58-E004+ & 1.57-E004*** \\
			& (1.45) & (4.96) & (1.90) & (4.64) \\
			\multirow{2}{*}{\textit{Suicide}} & 0.992 & 1.002 & -0.00597 & -0.00118 \\
			& (-0.83) & (0.37) & (-0.63) & (-0.21) \\
			\multirow{2}{*}{\textit{PC\_Income}} & 1.000 & 1.000* & -1.39-E005+ & -1.62E-005* \\
			& (-1.48) & (-2.18) & (-1.70) & (-2.14) \\
			\multirow{2}{*}{\textit{Urban}} &  & 1.004 &  & 0.00351 \\
			&  & (1.19) &  & (0.94) \\
			\multirow{2}{*}{\textit{Const.}} &  &  & 1.955** & 1.503** \\
			&  &  & (2.72) & (2.66) \\ \hline\hline
			\multicolumn{5}{l}{\footnotesize{t-statistics in parentheses. P-value: + p \textless{}0.1, * p \textless{}0.05, ** p \textless{}0.01, *** p \textless{}0.001.}}
	\end{tabular}}
\end{table}   
\section{Results and Discussion}    
In the presentation of results, we will focus solely on the Poisson regressions, given the difficult interpretation of the coefficients of the linear models discussed earlier. Note that all the reported coefficients relative to the Poisson models are incidence rate ratios (IRR). From Table \ref{tab2}, it is possible to observe that the Giffords score (lagged) is always statistically significant, although only at 10\% level in the RE models, while at 5\% in the FE models. An increase of one point in the overall Giffords score causes an approximately 1\% reduction in the number of fatal police shootings per million of inhabitants if considering the FE model. The percentage reduction is slightly lower for the RE model: 0.8\% when including all controls -- model (3) -- and 0.7\% when excluding the subset of jointly non-significant covariates -- model (4).\par 
An important point to notice is the lack of significance, in all models of Table \ref{tab2}, of the proxy for firearms diffusion: \textit{Suicide}. The strong correlation with the lag of the Giffords score ($\rho = -0.83$) may suggest that the effect of this last masks the one of the former. However, if running the same regressions without the inclusion of the Giffords score, the p--value of \textit{Suicide} remains above 0.1 in all models: 0.152, 0.106, 0.140 and 0.118 for, respectively, models (1), (2), (3) and (4) without \textit{Giff\_Score\_L} (full results available from the authors upon request). \par 
When considering Table \ref{tab3} and the disentangled categories composing the Giffords score, only the coefficient related to the lag of one category, gun owner accountability, is significant (at 5\% level). This happens to be true both in the FE and in the RE Poisson regressions, so as in the linear models. In particular, an increase of one point in the strength of the gun owner accountability category is associated with a decrease in per--million inhabitants fatal police shootings of 3.7\% (FE model) or of 3.8\% (RE model).\par 
We can further notice that in all models, both in Table \ref{tab2} and \ref{tab3}, the sign of the coefficients of the main control variables is as expected. In particular, the number of violent crimes positively impacts the number of fatal police shooting episodes whereas per--capita income has the opposite effect. A last word is dedicated to the significance of the violent crime rate that is very high (0.1\%) in all RE Poisson models but absent in the FE models. This is possibly due to the persistent nature in time of this phenomenon that, in the FE model, gets captured by the fixed effect component.           
\subsection{Discussion}
The present study shows that increasing levels of firearms regulation are significantly associated with a lower number of fatal police shooting cases. In particular, a point increase in the overall Giffords score leads to a decrease of in fatal police shootings of 0.7\%--1\%, depending from the model. When considering separately the various categories composing the Giffords score, one point increase in the strength of gun owner accountability leads to a decrease of approximately 3.7\%--3.8\% in the number of people killed by police officers. The diffusion of fire--weapons, instead, has no statistically significant role in determining the considered outcome. This finding has been achieved through the use of a panel dataset, thus controlling for unobserved heterogeneity through the use of FE Poisson models.\par
It is interesting to compare our results with previous findings. Regarding the effect of firearms diffusion, our results clearly contradicts the previous findings of \citet{kivisto17} and of \citet{hemenway19}. In fact, we do not find a statistically significant effect of guns diffusion in determining the number of fatal police shootings. \par 
Considering the strength of firearms regulations, our analysis basically confirms the findings of \citet{kivisto17}. However, this is true only for the overall score. When evaluating each category separately, significant differences emerge. First of all, it must be noted that the results are not easily comparable, given the different scoring system used in \citet{kivisto17}, namely the Brady scorecards, and in the present analysis. However, a comparison is not impossible. In \citet{kivisto17}, two categories remained significant after all controls were added, namely promoting safe storage via child and consumer safety laws and curbing gun trafficking. The former corresponds to the category consumers and child safety (\textit{CSS}) in the Giffords scorecards and the latter is included in other regulation of sales and transfers (\textit{ORST}), both not significant in our models. The gun owner accountability, the category found significant in the present analysis, is instead composed by three elements: licensing of gun owners and purchasers, having the highest weights, followed by registration of firearms and reporting lost or stolen firearms. This is an important difference, with potentially relevant implications for policy--makers. Furthermore, it is reasonable to think that the first sub-category, namely the need of gun owners to have a license, has a great discriminant power in determining the final identity of gun owners. This further suggests that the qualitative side of gun diffusion (who owns a gun) is more important in limiting the number of police shootings than the quantitative side (how many guns are owned).     
\section*{Conclusions}
The present analysis has shown that police shooting episodes are significantly reduced by stricter levels of firearms law. While this finding partially confirms what emerged in previous studies, we also find that the diffusion of fire--weapons is inconsequential in determining the number of police shootings, thus contradicting the precedent evidence.\par 
The policy recommendations that can be derived from the present paper are quite straightforward. Improving the strength of firearms regulations seems an effective way for reducing the number of people killed by law enforcement officers. Actually, the policy prescriptions can be even more specific. In fact, from the analysis it emerges that the most effective intervention for reducing fatal police shooting episodes would be to strengthen the rules of gun owner accountability, namely licensing of gun owners and purchasers, registration of firearms and reporting lost or stolen firearms. These policy prescriptions are different from the ones provided by previous studies.\par
Another important lesson, and a departure from previous findings, is the lack of statistical significance of the diffusion of guns in causing fatal police shootings. This suggests that the cause may be more qualitative (who owns the guns) rather than quantitative (how many guns). The fact that the only significant legislative category emerged from this study is the gun owner accountability further strengthens this hypothesis.\par  
There are several possible ways in which the analysis could be expanded in order to have more precise and specific prescriptions. One possibility would be to further disaggregate each category of the Giffords score into its subcategories. We have not pursued this road due to the limited number of observations at our disposal. Another interesting extension would be to consider the episodes of police shootings directed towards unarmed citizens \citep{hemenway19}. The lack of this information in the Fatal Encounters database has prevented us to conduct this analysis.    
\newpage 
\bibliographystyle{apalike}
\bibliography{Biblio}
\newpage
\section*{Appendix}\label{app}
\setcounter{figure}{0}
\renewcommand{\thefigure}{A\arabic{figure}}
\begin{figure}[htbp!]
	\centering
	\caption{Distribution of the Dependent Variable (\textit{Pol\_shoot}) before and after the Yeo-Johnson Power Transformation ($\lambda = 0.1384$)}\label{figs1}
	\begin{subfigure}{.7\textwidth}
		\centering
		\caption{Before}
		\includegraphics[width=0.95\linewidth]{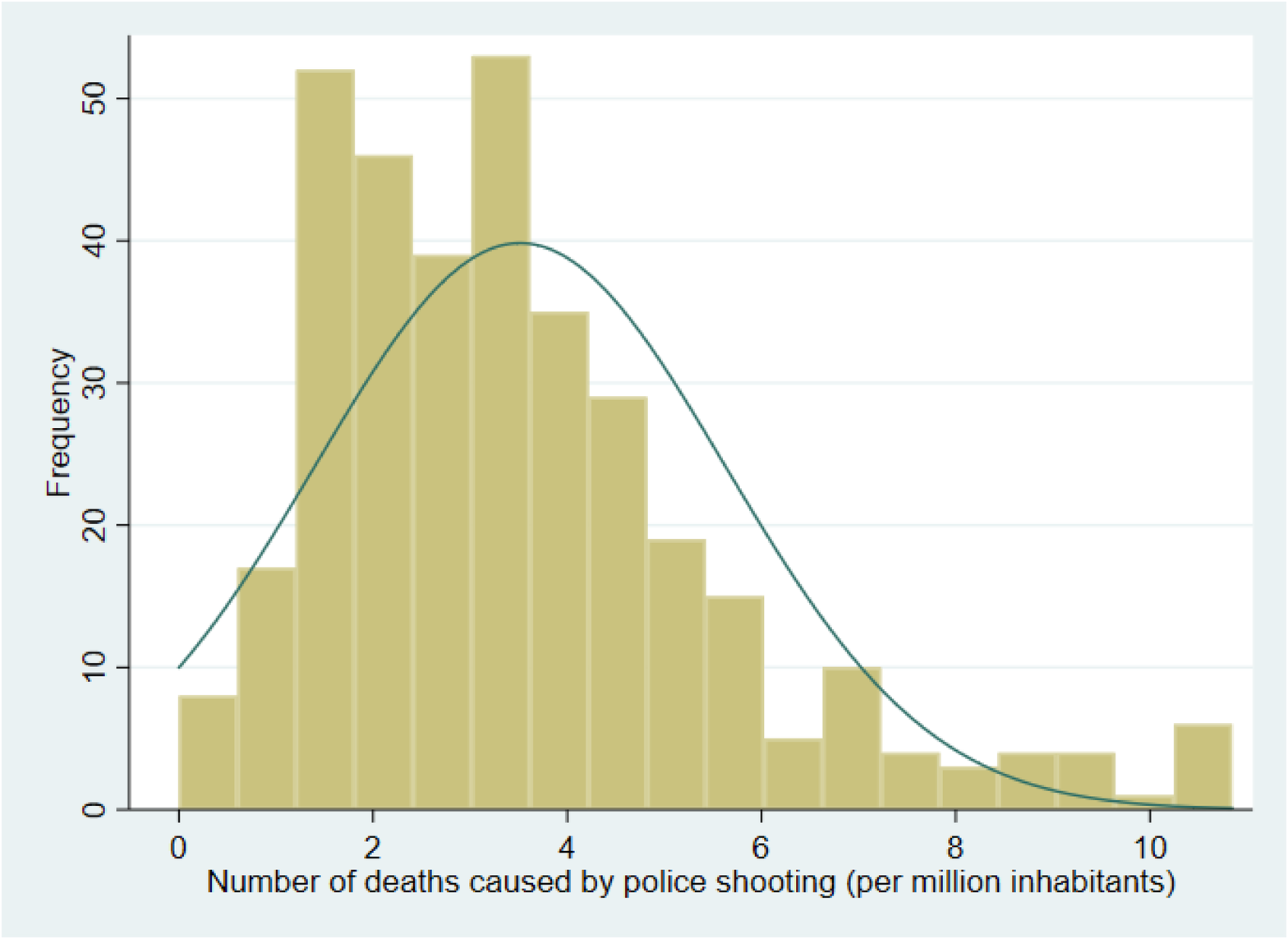}
	\end{subfigure}\\\vspace{10pt}
	\begin{subfigure}{.7\textwidth}
		\centering
		\caption{After}
		\includegraphics[width=0.95\linewidth]{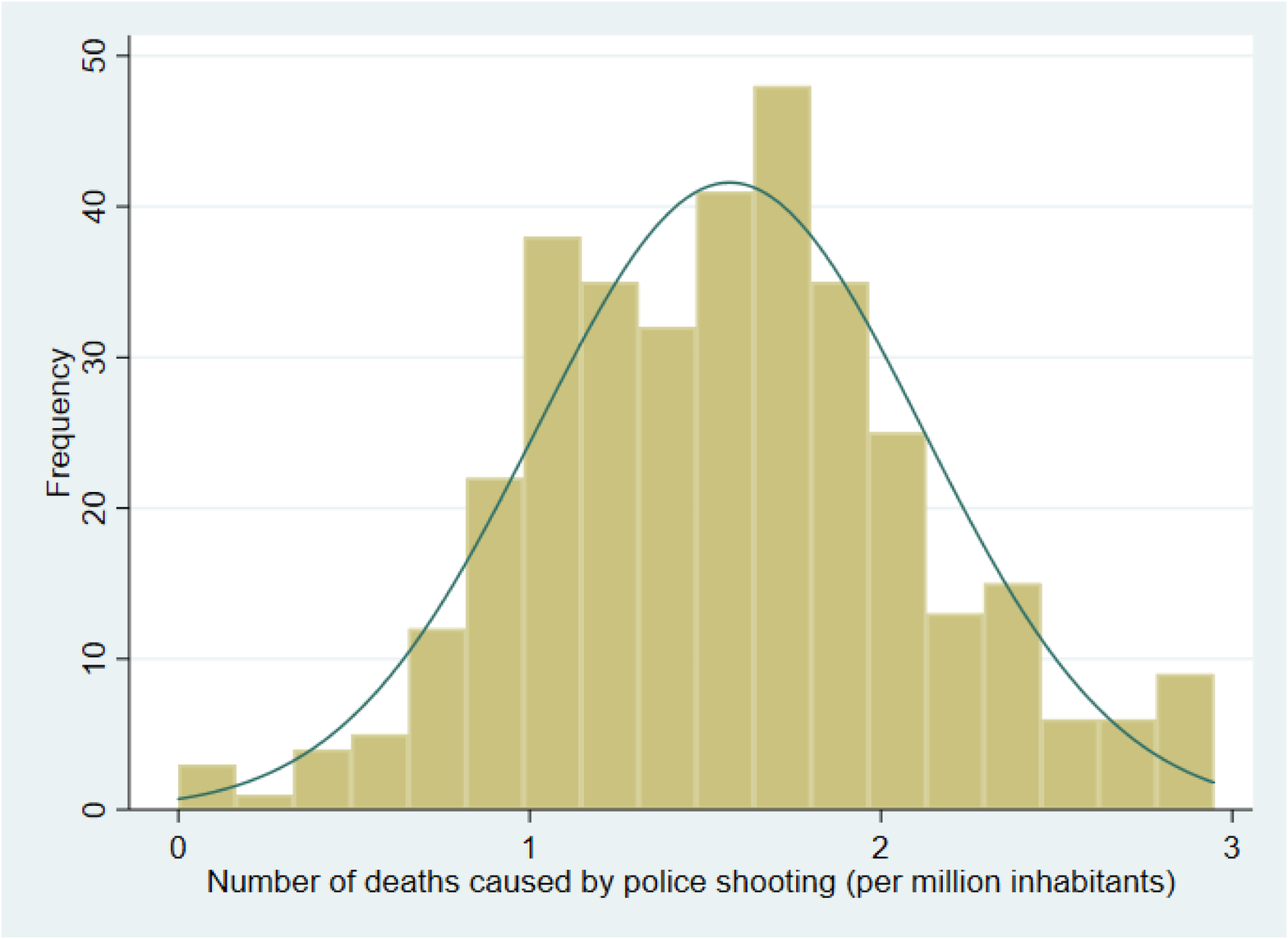}
	\end{subfigure}
\end{figure}

\setcounter{table}{0}
\renewcommand{\thetable}{A\arabic{table}}
\begin{table}[htbp!]
		\centering
	\caption{Ancillary Information for Table \ref{tab2} and Table \ref{tab3}}\label{tabs1}
%	\resizebox{0.8\textwidth}{!}{
	{\renewcommand{\arraystretch}{1.3}
	\begin{tabular}{lcccc}
		\hline\hline
		\multicolumn{5}{c}{\textit{\textbf{Ancillary Information for Table \ref{tab2}}}} \\
		 & (1) & (2) & (3) & (4) \\
		\textit{2014} & 1.089 & 1.105 & 1.055 & 1.081 \\
		\textit{2015} & 1.106 & 1.137 & 1.071 & 1.102 \\
		\textit{2016} & 1.149 & 1.190* & 1.137* & 1.148** \\
		\textit{2017} & 1.224 & 1.285* & 1.193* & 1.220** \\
		\textit{2018} & 1.363** & 1.439** & 1.308*** & 1.344*** \\
		\textit{ln($\alpha$)} &  & 0.0540 &  & 0.0545 \\
		\textit{RESET Test}$^1$ & 0.316 & 0.864 & 0.859 & 0.941 \\
		\textit{Mundlak Test}$^2$ & \multicolumn{2}{c}{0.0555} & \multicolumn{2}{c}{0.3221} \\
		\textit{N. Obs.} & 300 & 300 & 300 & 300 \\ \hline
		\multicolumn{5}{c}{\textit{\textbf{Ancillary Information for Table \ref{tab3}}}} \\
		 & (5) & (6) & (7) & (8) \\
		\textit{2014} & 1.164 & 1.137 & 0.129 & 0.0943 \\
		\textit{2015} & 1.173 & 1.152 & 0.128 & 0.0985 \\
		\textit{2016} & 1.234** & 1.199*** & 0.188** & 0.159** \\
		\textit{2017} & 1.288** & 1.264*** & 0.199* & 0.176** \\
		\textit{2018} & 1.411*** & 1.397*** & 0.286** & 0.265*** \\
		\textit{ln($\alpha$)} &  & 0.0463 &  &  \\
		\textit{RESET Test}$^1$ & 0.405 & 0.976 & 0.880 & 0.659 \\
		\textit{Mundlak Test}$^2$ & \multicolumn{2}{c}{0.3885} & \multicolumn{2}{c}{0.2854} \\
		\textit{N. Obs.} & 300 & 300 & 300 & 300\\\hline\hline
	\end{tabular}}\\
\begin{adjustwidth}{0.8in}{0.8in}
\scriptsize{2013 as base year}.\\
\scriptsize{1) The RESET Test reports the p--value of the squared residuals.}\\
\scriptsize{2) The Mundlak Test reports the p-value of the joint significance test of the time average of all regressors.}
\end{adjustwidth}
\end{table}
\end{document}